\documentclass[preprint,showpacs,preprintnumbers,amsmath,amssymb]{revtex4}
\usepackage{graphicx}
\usepackage{dcolumn}
\usepackage{bm}
\begin{document}
\title{The Parallelometer: a mechanical device to study curvature }
\title{The Parallelometer: a mechanical device to study curvature }
  \author{Alberto G. Rojo}
  \email{rojo@oakland.edu}
  \author{David Garfinkle}
  \email{garfinkl@oakland.edu}
 \affiliation{%
Department of Physics, Oakland University, Rochester, MI 48309}
\date{\today}


\begin{abstract}
A simple mechanical device is introduced, the parallelometer, that
can be used to measure curvatures of surfaces. The device can be
used as a practical illustration of parallel transport of a vector.
Its connection to the Foucault pendulum is discussed.

\end{abstract}
\maketitle
\section{Introduction}

It is intuitively obvious how to transport a vector in a plane in
such a way that it remains unchanged: just pick a set of cartesian
coordinates and ask that the components of the vector   remain
constant as it is transported. As it is transported, the vector
remains parallel to its initial direction in the plane. How about in
a curved surface? This problem was first addressed by Levi-Civita in
1917\cite{levy} who introduced the influential idea of parallel
transport, later to be used extensively in the General Theory of
Relativity.  To parallel transport a vector along a path on a
(two--dimensional) curved surface, we can first refer it to a local
cartesian system which in turn has to be parallel transported. The
$z$ axis is always normal to the surface. Qualitatively, the
prescription of transport is that, as one moves along the curve,
the $xy$ plane cannot ``rotate" around the $z$ axis. A vector will
be parallel transported when its components with respect to these
$xy$ axes don't change.  A non-trivial result emerges: if the vector
is parallel-transported around a closed curve, it returns rotated
with respect to the initial orientation.

In this paper we present
a simple mechanical device that measures parallel transport and can
be used as a pedagogical tool. Here the phase factor in the parallel
transport becomes very clear and easy to observe and study
quantitatively. In principle an experiment should be easy and
inexpensive and can be carried out in a basic laboratory setting,
adequate for beginner or intermediate students. The system we
devised, the parallelometer, consists in general  of two concentric
rings constrained to rotate without friction in the same plane.
Consider the inner ring carried along a path maintaining its plane
tangent to the surface. Since the outer ring is free to rotate, a
vector connecting the center of the parallelometer to a point in the
outer ring will be parallel transported along the path.\cite{bicycle} The
practical realizations could be a ball bearing or a freewheel.


\section{The Parallelometer}

\subsection{ Lagrangian treatment}
\label{lagrange}



\begin{figure}[h]
\vspace{0.cm}
\begin{center}
\includegraphics[width=0.55\textwidth]{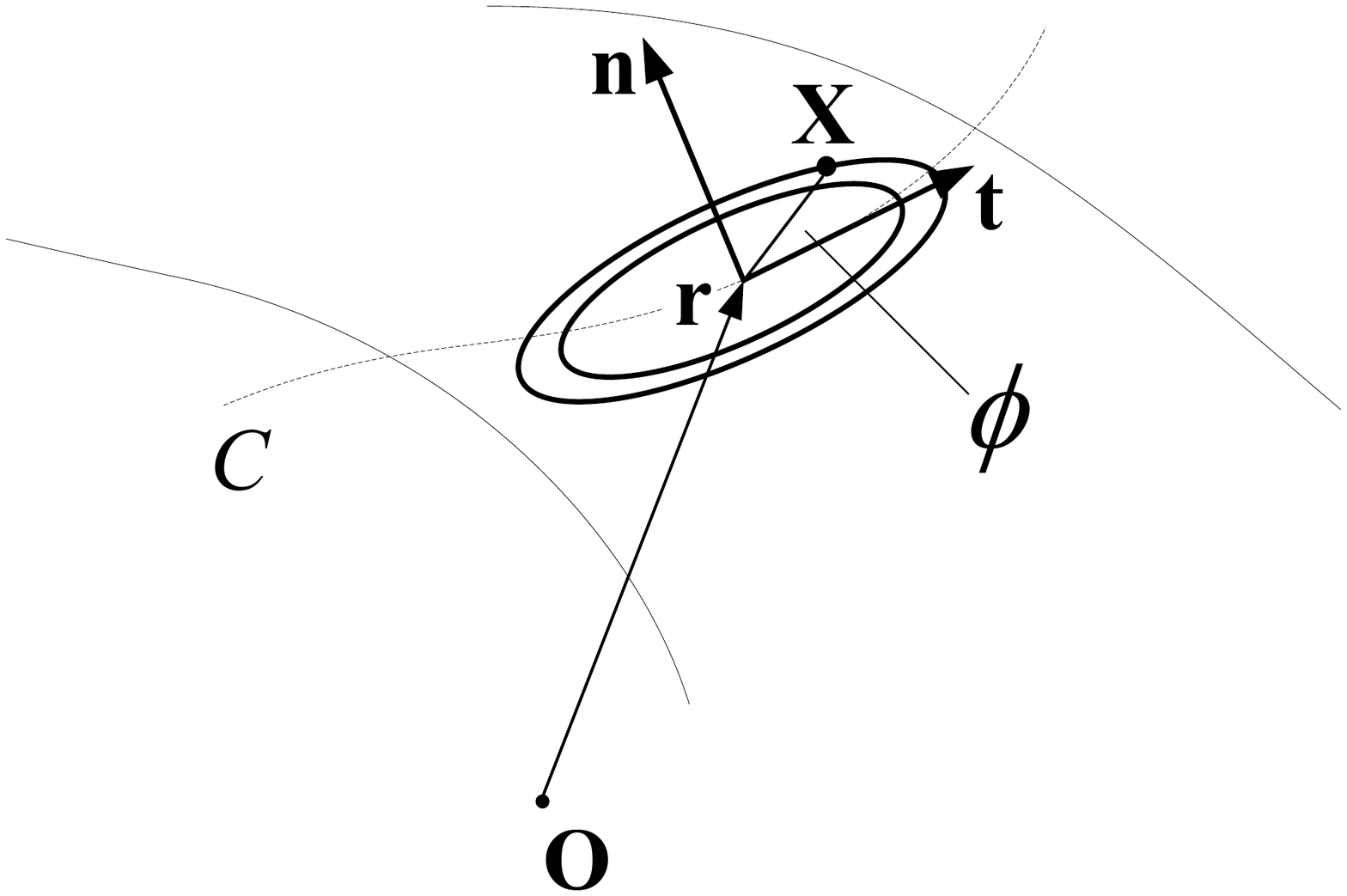}
\vspace{-6.cm}
\end{center}
\caption{The parallelometer consists of a frictionless freewheel
which is restricted to move along a given surface with its axis
pointing along the local normal to the surface.  This device will be
transported along a curve $C$ on a surface $S$. }
 \label{angmray}
\end{figure}

In this section we  discuss the system dynamics and show that it
corresponds to parallel transport. The parallelometer, consists of a
frictionless freewheel which is restricted to move along a given
surface with its axis pointing along the local normal to the surface
This device will be transported along a curve $C$ on a surface $S$.
(See Fig. 1).
In this process the parallelometer is constrained to move with its
plane always tangent to the surface and to move along the curve $C$.
%
For concreteness we can think of  an arrow attached to the axis of
the freewheel (and perpendicular to the axis) that always points
along the curve $C$.



The coordinate of a point ${\textbf{X}}$ within the wheel, of radius
$r_0$, is

\begin{equation}
\textbf{X}=\textbf{r}+r_0\left[\cos \beta {\textbf{t}}+ \sin \beta
(\textbf{n}\times {\textbf{t}})\right],
\end{equation}
where ${\textbf{t}}$ is a unit vector in the direction of  the local
tangent to the curve $C$ along which the parallelometer is
transported. The unit vector $\textbf{n}$ is the local normal to the
plane tangent to the surface, and $\textbf{r}$ is the coordinate of
the center of the wheel. Notice that, with these definitions, the
unit vectors ${\textbf{t}}$ and $\textbf{n}\times {\textbf{t}}$ are
along the natural $x$ and $y$ axis for an observer on an ``inner
ring" defined as the intersection of the (cylindrical) axis of the
freewheel with the tangent plane. The vector $\textbf{n}$ is along
the corresponding $z$ axis.  This frame is similar to the standard Frenet
frame associated with a curve that is often used in differential
geometry.\cite{spivak}  However, it is not identical to a Frenet frame, since
its construction depends not on the curve alone but also on the surface.

The velocity of that point on the free wheel is:

\begin{eqnarray}
\dot{\textbf{X}}&=&\dot{\textbf{r}}+r_0\bigl\{\dot{\beta}\left[-\sin
\beta {\textbf{t}}+ \cos \beta (\textbf{n}\times
{\textbf{t}})\right] + \cos \beta \dot{{\textbf{t}}}
\nonumber
\\
&+& \sin \beta
(\dot{{\textbf{n}}}\times {\textbf{t}}+ \textbf{n}\times
\dot{{\textbf{t}}})\bigr\}
\end{eqnarray}

Squaring $\dot{\textbf{X}}$ and integrating over $\beta$ we obtain
the kinetic energy of the ring. The Lagrangian is therefore
\begin{equation}
L={\frac M {2 \pi}} \int_0^{2\pi} d\beta\;\dot{\textbf{X}}^2
\end{equation}

Using
\begin{eqnarray} \int_0^{2\pi} d\beta\, \sin\beta \cos \beta=0,
\nonumber
\\
\int_0^{2\pi} d\beta\, \sin ^2 \beta=\int_0^{2\pi} d\beta\,
\cos ^2\beta =\pi
\end{eqnarray}
we obtain the corresponding Lagrangian:
\begin{equation}
L= {\frac 1 2}M\dot{\textbf{r}}^2 +
{\frac 1 2}Mr_0^2\dot{\beta}^2+Mr_0^2\left[
\dot{{\textbf{t}}}\cdot(\textbf{n}\times
{{\textbf{t}}})\,\dot{\beta}\right]+L'
\end{equation}
where $L'$ is a term that does not contain  $\beta$ or $\dot{\beta}$
and therefore does not affect the dynamics of the angle $\beta$.
Notice that the term $r_0 \dot{{\textbf{t}}}\cdot(\textbf{n}\times
{{\textbf{t}}})$ plays the role of a ``gauge potential"  affecting
the motion of $\beta $.

 The equation of motion of $\beta$ is
\begin{equation}
{\frac d {dt}}{\frac {\partial L} {\partial \dot {\beta}}}
= {\frac {\partial L}
{\partial \beta}},
\end{equation}
and, since $L$ is independent of $\beta$ we have
\begin{equation}
{\frac {\partial L} {\partial \dot{\beta}}}=\rm{Const},
\end{equation}
or
\begin{equation}
 \dot{\beta}+ \dot{{\textbf{t}}}\cdot(\textbf{n}\times
{{\textbf{t}}})=\rm{Const}.
\end{equation}

If, as an initial condition we choose $\dot{\beta}=0$ and
$\dot{{\textbf{t}}}=0$  then Const$=0$ and we obtain

\begin{equation}
 \dot{\beta}+ \dot{{\textbf{t}}}\cdot(\textbf{n}\times
{{\textbf{t}}})=0, \label{dyn}
\end{equation}

We will now show that this dynamics corresponds to parallel
transport {\em{provided}} the initial condition corresponds to
 Const$=0$.

{\subsection{Connection with parallel transport}} \label{parallel}

Consider now transporting a unit vector ${\textbf{u}}$ along the
same curve $C$. Following the previous discussion, we can define a
local coordinate system with the unit vectors $(\textbf{t},
\textbf{n}\times {\textbf{t}}, \textbf{n})$.
Notice that this a cartesian system natural for an observer that
faces along the direction of the motion on the curve and
{\em{not}} the coordinate system that is parallel transported.
 Since the vector is always in the plane
tangent to the surface we need only one coordinate to specify it
in this coordinate system. We use the  angle $\beta$ that  the
vector makes with the tangent  $\textbf{t}$ to the curve. This
means that ${\textbf{u}}$ can be expressed in the following way in
terms of the local coordinates:
\begin{equation}
{{\textbf{u}}}=\cos \beta {\textbf{t}}+ \sin \beta
(\textbf{n}\times {\textbf{t}}). \label{vec1}
\end{equation}

 Now multiply the above equation by ${\textbf{t}}$ and, using
${\textbf{t}}\cdot(\textbf{n}\times {\textbf{t}})=0$, compute the
time derivative. Since \begin{equation}
{\textbf{t}}\cdot\textbf{u}=\cos \beta, \end{equation} we have
\begin{eqnarray}
-\dot{\beta}\sin\beta&=&{\frac d {dt}}{\textbf{u}}\cdot
{\textbf{t}}\nonumber \\
&=& \dot{{\textbf{u}}}\cdot {\textbf{t}} + {{\textbf{u}}}\cdot
\dot{{\textbf{t}}} \label{115}
\end{eqnarray}
For parallel transport of a vector two conditions have to be
satisfied. First, the vector has to stay in the plane tangent to
the surface. This condition is guaranteed by the form of Eq.
(\ref{vec1}). The second condition is that any variation of the
vector transported is out of the plane. These two conditions
combined contain the prescription for how to parallel transport a
vector.  The second condition in this case means \begin{equation}
\dot{{\textbf{u}}}\cdot {\textbf{t}}=0,
\end{equation}
so that, from (\ref{115})
\begin{equation}
-\dot{\beta}\sin\beta={{\textbf{u}}}\cdot \dot{{\textbf{t}}}.
\end{equation}
Replacing (\ref{vec1}) in the above equation, and using $
{\textbf{t}}\cdot \dot{{\textbf{t}}}=0$ (the variation of a vector
of constant length is always perpendicular to the vector) we obtain
\begin{equation}
-\dot{\beta}=\dot{{\textbf{t}}}\cdot(\textbf{n}\times
{{\textbf{t}}}) \label{114}
\end{equation}
which is the same as (\ref{dyn}), obtained from dynamical
considerations. The parallelometer therefore measures parallel
transport provided it is started at rest.

{\subsection{Connection with Foucault's pendulum}}

Consider the above results specified to a spherical surface and the
curve $C$ being a parallel. In spherical coordinates, where $\theta
$ is the polar angle and $\phi$ the asymuthal angle, we have
\begin{equation}{\textbf{t}} =\textbf{e}_\phi,
\end{equation}
\begin{equation}
{\dot{{\textbf{t}}}} =-\left(\sin\phi \hat{\textbf{i}}+\cos\phi
\hat{\textbf{j}}\right)\dot{\phi},
\end{equation}
and
\begin{equation}
\textbf{n}\times {{\textbf{t}}}=-\cos\theta \left(\sin\phi
\hat{\textbf{i}}+\cos\phi \hat{\textbf{j}}\right)+\sin\theta
\hat{\textbf{k}},
\end{equation}
where $\theta=\pi/2 -\lambda $ is complementary to the latitude
$\lambda$. This implies, from (\ref{114})
\begin{equation}
\dot{\beta}=-\cos \theta \dot{\phi}, \label{foucault2}
\end{equation}

or equivalently

\begin{equation}
d{\beta}=-\cos \theta d{\phi}. \label{foucault3}
\end{equation}

Integrating over $2\pi$ we obtain

\begin{equation}
\Delta{\beta}\equiv \delta=-2\pi \cos \theta. \label{foucault4}
\end{equation}

which is the
equation for Foucault's pendulum.

\section*{Acknowledgements}

A. G. R. thanks the Research Corporation for
support. D.G. was supported by NSF grant PHY-0456655.

\end{document}